\documentclass{article}

\usepackage{main}
\usepackage[utf8]{inputenc} 
\usepackage[T1]{fontenc}    
\usepackage{hyperref}       
\usepackage{url}            
\usepackage{booktabs}       
\usepackage{amsfonts}       
\usepackage{nicefrac}       
\usepackage{microtype}      
\usepackage{graphicx}
\usepackage{doi}
\usepackage{tabularx}
\newcolumntype{C}{>{\centering\arraybackslash}X}

\title{Modelling Moral Traits with Music Listening Preferences and Demographics}

\author{ 
	Vjosa Preniqi\\
	Centre for Digital Music \\
	Queen Mary University of London\\
	\texttt{vjosa.preniqi@qmul.ac.uk} \\
	\And
	Kyriaki Kalimeri \\
	ISI Foundation\\
	\texttt{kyriaki.kalimeri@isi.it} \\
	\And
	Charalampos Saitis \\
	Centre for Digital Music \\
	Queen Mary University of London\\
	\texttt{c.saitis@qmul.ac.uk} \\
}

\renewcommand{\shorttitle}{Music and Moral Traits}

\hypersetup{
pdftitle={Music and Moral Traits},
pdfauthor={Vjosa Preniqi, Kyriaki Kalimeri, Charalampos Saitis},
pdfkeywords={Moral values, Music genres, Musical Preferences},
}

\begin{document}
\maketitle

\begin{abstract}
Music is an essential component in our everyday lives and experiences, as it is a way that we use to express our feelings, emotions and cultures. In this study, we explore the association between music genre preferences, demographics and moral values by exploring self-reported data from an online survey administered in Canada. Participants filled in the moral foundations questionnaire, while they also provided their basic demographic information, and music preferences.
Here, we predict the moral values of the participants inferring on their musical preferences employing classification and regression techniques.
We also explored the predictive power of features estimated from factor analysis on the music genres, as well as the generalist/specialist (GS) score for revealing the diversity of musical choices for each user. Our results show the importance of music in predicting a person's moral values (.55-.69 AUROC); while knowledge of basic demographic features such as age and gender is enough to increase the performance (.58-.71 AUROC). 
\end{abstract}


\section{Introduction}

Music influences a wide range of behaviours including cognitive performance such as thinking, reasoning, problem-solving, creativity, and mental flexibility \cite{macdonald2013music}, while it is proven to have therapeutic effects in various medical situations \cite{lin2011mental}.
Scientists showed that music is a fundamental factor in the evolution of humans and societies, and it is tightly related to social connection, communication and development \cite{cross2009evolution,loersch2013unraveling}.
Music is shown to improve concentration while working, boost the stimulation during the fitness workout, regulate the moods, and it helps to relieve stress and anxiety \cite{north2004uses,lonsdale2011we}. 
A notable influence on increasing rehabilitation rates after surgery \cite{hole2015music}, and capturing mental health problems such as depression, autism, post-traumatic disorders and so on \cite{carr2012group,gold2009dose} was also associated with music.

More recently, evidence that music listening behaviours reflect on people’s personality traits and values 
were provided analysing on a variety of data including self-reported surveys~\cite{devenport2019predicting,gardikiotis2012rock,greenberg2015personality}, online music streaming~\cite{anderson2020just,krismayer2019predicting}, and social media~\cite{nave2018musical}. Findings from surveys have showed that musical taste is strongly related to personal values \cite{gardikiotis2012rock} and political orientation \cite{devenport2019predicting} and that personality can predict musical sophistication even after controlling for demographic variables and musicianship \cite{greenberg2015personality}. 
Using data from the popular but controversial myPersonality Facebook application, Nave et al. \cite{nave2018musical} found that both people's reactions to unfamiliar music samples and ``likes'' for music artists predicted personality traits. 
A study on the social music platform Last.fm showed that the music listening behaviours of users can predict their demographics, including age, gender, and nationality \cite{krismayer2019predicting}.
Anderson et al. \cite{anderson2020just} presented strong evidence about the connection between personalities and music listening preferences by utilising Spotify music streaming data. 
Such knowledge has the potential to not only better our understanding of the social psychology of music, but also to further support the improvement of personalised content and enhance data-driven decision-making in music marketing and entertainment. 

Building on comparable interactionist theories, here we set to explore the less attended relation between moral values and music preferences. 
We operationalise morality according to the Moral Foundations Theory (MFT)\cite{Graham2011}, which defines five dimensions of morality, namely \textbf{Care/Harm, Fairness/Cheating, Loyalty/Betrayal, Authority/Subversion}, and \textbf{Purity/Degradation}. These can further collapse into two superior moral foundations: of \textbf{Individualising}, compounded by fairness and care,  that asserts that the basic constructs of society are the individuals and hence focuses on their protection and fair treatment, and of \textbf{Binding}, that summarises purity, authority and loyalty and is based on the respect of leadership and traditions. Moral foundations are considered to be higher psychological constructs than the more commonly investigated personality traits \cite{mcadams2006new} yet the moral values have attracted less attention from music scientists.

In the computational social science field, moral foundations have been studied only recently. Kim and colleagues \cite{kim2013moral} observed digital traces aggregated from repeated game play data in order and pointed out several associations with people’s values and moral identity. Kalimeri et al. \cite{kalimeri2019human} tried to predict moral foundations from digital behaviour traits collected from users' smartphones and browser activities. Their results showed that moral traits and values are more complex, and thus harder to predict compared to demographics which can be more accurately inferred from passively collected data. Nevertheless, these studies provide a realistic dimension of the possibilities of modelling moral traits for delivering better targeted and more effective interventions. Indeed, recent work has indicated that negative emotions enforced by types of music can worsen moral judgement \cite{ansani2019you} although that study did not rely on a psychometrically validated tool like the MFT.

\section{Data Collection and Feature Engineering}
\label{sec:headings}

Here, we employ data from a survey administered online for a general scope marketing project. The survey consists of approximately 2,000 participants (51\% females) from 12 different regions in Canada. The participants filled in, among other items, information about their demographics, including age, gender, education, and political views. They completed the validated Moral Foundations questionnaire~\cite{haidt2004intuitive}, while stated their preferences on 13 music genres (on a 5-point Likert scale where 1 = strongly dislike and 5 = strongly like). The considered music genres were: alternative pop/rock, christian, classical, country, folk, heavy metal, rap/hip-hop, jazz, latin, pop, punk, R\&B, and rock. 
After an initial pre-processing step, during which we removed participants who failed in the catch items of the survey, we remained with 1,062 participants
(55\% females). Table ~\ref{tab:Canada_ds_demographics} summarises the major demographic features of our dataset.


In addition to using the ratings on each of the 13 music genres, we considered two steps of dimensionality reduction as part of feature engineering. First, to reveal the underlying structure of the 13 genre preferences, we performed an exploratory factor analysis using principal axis factoring with promax rotation. A 5-factor solution was retained, which explained 67\% of total data variance: \{jazz, classical, latin\}, \{punk, heavy metal, rap/hip-hop\}, \{pop, R\&B\}, \{country, christian, folk\}, and \{rock, alternative pop/rock\} (genres ordered by decreasing factor loading). Similar factors have been reported for Greek listeners \cite{gardikiotis2012rock}. 
Second, to quantify the musical diversity of a given respondent, the loadings of each of the 13 music genres onto the five factors were considered as that genre's vector representation in the ``preference space'' and were used to compute a type of generalist-specialist (GS) score, inspired by the work of Anderson et al. \cite{anderson2020algorithmic}. Intuitively, generalists versus specialists will have genre vectors spread apart versus close together in the preference space.  
We first calculate the user centroid $\overrightarrow{ct_i}$ of the genre vectors representing the loadings of the genre on the five factors $\overrightarrow{l_j}$, weighted by the number of genre scores rated by each respondent $w_j$: 

\begin{equation}
  \overrightarrow{ct_i} = \frac{1}{\sum{w_j}} \cdot \sum{w_j \overrightarrow{l_j}} \;  .
\end{equation}
The GS score is then the average cosine similarity between a genre vector and the preference-weighted average of a participants' genre vectors: 

\begin{equation}
    GS(u_i) = \frac{1}{\sum{w_j}} \cdot \sum w_j \frac{\overrightarrow{l_j} \cdot \overrightarrow{ct_i}}{\parallel{\overrightarrow{l_j}\parallel} \cdot \parallel{\overrightarrow{ct_i}}\parallel } \;  .
\end{equation}

\begin{table}[ht]
\centering
\caption{The summary of our dataset (cleaned) with major demographic attributes utilised for this research work. These labels were used as predictors for MFT categories.}
\begin{tabularx}{\textwidth}{ @{}X X C@{} }
\toprule
Attributes & Demographics & Sample size (N = 1062)\\ 
\midrule
\textbf{Age}  &  18-24 & 80 (7.5\%) \\
& 25-34 & 154 (14.5\%) \\
& 35-44 & 205 (19.3\%) \\
& 45-54 & 205 (21.9\%) \\
& 55-64 & 187 (17.6\%) \\
& 65+ & 203 (19.1\%) \\
\midrule
\textbf{Gender} & Female & 588 (55.3\%) \\
& Male & 474 (44.6\%) \\
\midrule
\textbf{Education} & Less than High School & 35 (3.2\%) \\ 
& High school graduate & 195 (18.3\%) \\ 
& Some College  & 154 (14.5\%) \\ 
& Trade or professional school  & 115 (10.8\%) \\ 
& College Graduate  & 349 (32.8\%) \\  
& Post Graduate work or degree  & 205 (19.3\%) \\
\midrule
\textbf{Political Party} &  Conservative &  328 (30.8\%) \\ 
& Liberal &  279 (26.2\%) \\ 
& New Democratic Party &  184 (17.3\%) \\ 
& Green Party &  66 (6.2\%) \\ 
& Party Quebecois &  56 (5.2\%) \\ 
& I don't vote  &  149 (14\%) \\ 
\bottomrule
\end{tabularx}
\label{tab:Canada_ds_demographics}
\end{table}

\section{Experiments and Results}

\subsection{Exploratory Analysis}

We explored the extent to which musical genres’ preferences, demographics, political views and moral traits are correlated to each-other. We observed a positive Spearman correlation of age with christian,  classical,  country  and  folk  music  genres ($\rho_{s} = \{0.18$, 0.21, 0.20, 0.25\}),  respectively.
On the other side, heavy metal, hip-hop/rap and punk were more preferred by younger ages, whereas older people expressed their dislike towards these genres ($\rho_{s} = \{$-$0.22$,  $-$0.38,  $-$0.38 \}).

Education was positively related with classical music, jazz and latin music ($\rho_{s} = \{0.22$, 0.13, 0.13\}), indicating that people with higher education preferred these genres. With respect to morals, loyalty, authority and purity were positively correlated with christian music ($\rho_{s}=\{0.18$, 26, 38\}) and country music ($\rho_{s}=\{0.17$, 20, 21\}). 
Negative correlations were observed between purity and more rebellious and non-mainstream genres such as alternative pop/rock, heavy metal, punk  and rock ($\rho_{s}=\{$-$0.16$, $-$0.14,  $-$0.21, $-$0.16 \}). 
Looking at the political views of the respondents, conservatives were positively correlated with christian and country ($\rho_{s} =\{ 0.12$, 0.12\}) and negatively correlated to hip-hop/rap and punk ($\rho_{s} = \{$-$0.17$, $-$0.15\}). 


Correlating the GS score with moral foundations we observed a moderate negative correlation of the GS score and Purity,  ($\rho_{s}=$-$ 0.17$). Purity foundation is shaped by the psychology of disgust and contamination; hence, likely the participants who score high on this foundation, are translating these notions to the music domain, listening only one or few music genres, avoiding exposure to multiple - less ``noble'' - music genres.

\subsection{Moral Values Classification}

Our main research question is whether we can predict peoples' moral values from their music preferences. To answer this question, we postulate the task as a supervised classification one, developing a series of experiments to assess the predictive power of different variables (see Table~\ref{tab:experimental_design}).

We assign the class label ``high'' to individuals with moral scores higher than the population median for the specific foundation, and  ``low'', otherwise.
We perform 5-fold cross-validation on shuffled data (to avoid dependencies in successive data points), with 70\% of training and 30\% testing data.
We opt for the gradient boosting algorithm XGBoost (XGB) \cite{chen2016xgboost}.  
To take into account the effect of unbalanced class labels in the performance metric, we evaluate our models with the area under the receiver operating characteristic (AUROC) metric. The best model is then chosen as the one that maximized the weighted area under receiver operating characteristic (AUROC) statistic.

\begin{table}[ht]
\centering
\caption{Detailed list of the experiments we performed with the list of features employed as predictors in each one of them.}
\label{tab:experimental_design}
\begin{tabularx}{.7\textwidth}{>{\hsize=.25\hsize}C >{\hsize=.75\hsize}X}
\toprule
Experiment ID & Features Employed as Predictors \\
\midrule
EX1 & 13 Music Genres\\
EX2 & 5 factors\\
EX3 & GS score\\
EX4 & 13 Music Genres, Age, Gender\\
EX5 & 13 Music Genres, Age, Gender, Education\\
EX6 & 13 Music Genres, Age, Gender, Education, Political Views\\
\bottomrule
\end{tabularx}
\end{table}

\begin{table}[ht]
\centering
\caption{Moral traits classification with XGBoost, average weighted AUROC and standard deviation over 5-fold cross-validation (baseline is .50).}
\label{tab:MFT_predictions_xgb}
\begin{tabularx}{.7\textwidth}{@{}X C C C}
\toprule
\multicolumn{4}{c}{Music Preference - Classification Task} \\
\midrule
 & EX1 & EX2 & EX3 \\
\midrule
Care & \textbf{.57 (3.7)} & .54 (2.1) & .52 (1.5) \\
Fairness & \textbf{.56 (2.9)} & .52 (1.1) & .48 (2.7) \\
Authority & \textbf{.63 (0.8)} & .60 (1.1) & .49 (1.7) \\
Purity & \textbf{.69 (2.8)} & .65 (3.0) & .57 (2.3) \\
Loyalty & \textbf{.61 (2.4)} & .56 (1.9) & .48 (3.1) \\
\midrule
Individ. & \textbf{.55 (3.5)} & .51 (0.8) & .50 (1.6) \\
Binding & \textbf{.67 (2.4)} & .63 (2.2) & .52 (1.9)\\
\bottomrule
\end{tabularx}
\end{table}

Initially, we compared the predictive power of the genre information against the features engineered by us (EX1, EX2, and EX3). We trained one model per moral foundation, and we present the cross validated results in Table \ref{tab:MFT_predictions_xgb}.
We notice that the information obtained directly about the music preferences (EX1) outperforms the features we developed.
When comparing the scenarios, we observe that the five factors, and  the GS score accounting only for part of the variance in the data, did not manage to outperform the explicit information on music preferences.

A question that emerges naturally, is whether including knowledge regarding the participants' basic demographic features (i.e. age, political views, education level)  will improve the prediction of their moral values.
Table ~\ref{tab:MFT_predictions_xgb_2} summarises the results when age, gender, education and political views are incorporated in the design. As expected, the more information we have about the participants the more precise our predictions become, however, the improvement is minimum. This shows us the importance of music behaviours alone in explaining the variability of our moral values.


Further, we employed SHAP (SHapley Additive exPlanations), a game theory approach developed to explain the contribution of each feature to the final output of any machine learning model. 
SHAP values provide both global and local interpretability, meaning that we can assess both how much each predictor and each observation, respectively, contribute to the performance of the classifier.
The local explanations are based on assigning a numerical measure
of credit to each input feature. Then, global model insights can be obtained by combining many local explanations from the samples~\cite{lundberg2019explainable}.
As mentioned by the authors, the classic Shapley
values can be considered ``optimal'' in the sense that within a large class of approaches, they are the only way
to measure feature importance while maintaining several natural properties from cooperative game theory~\cite{lundberg2017unified}.
SHAP's output helps to understand the general behaviour of our model by assessing the impact of each input feature in the final decision, thus enhancing the usefulness of our framework.

\begin{figure}[ht]
    \centering
    \includegraphics[scale=0.55]{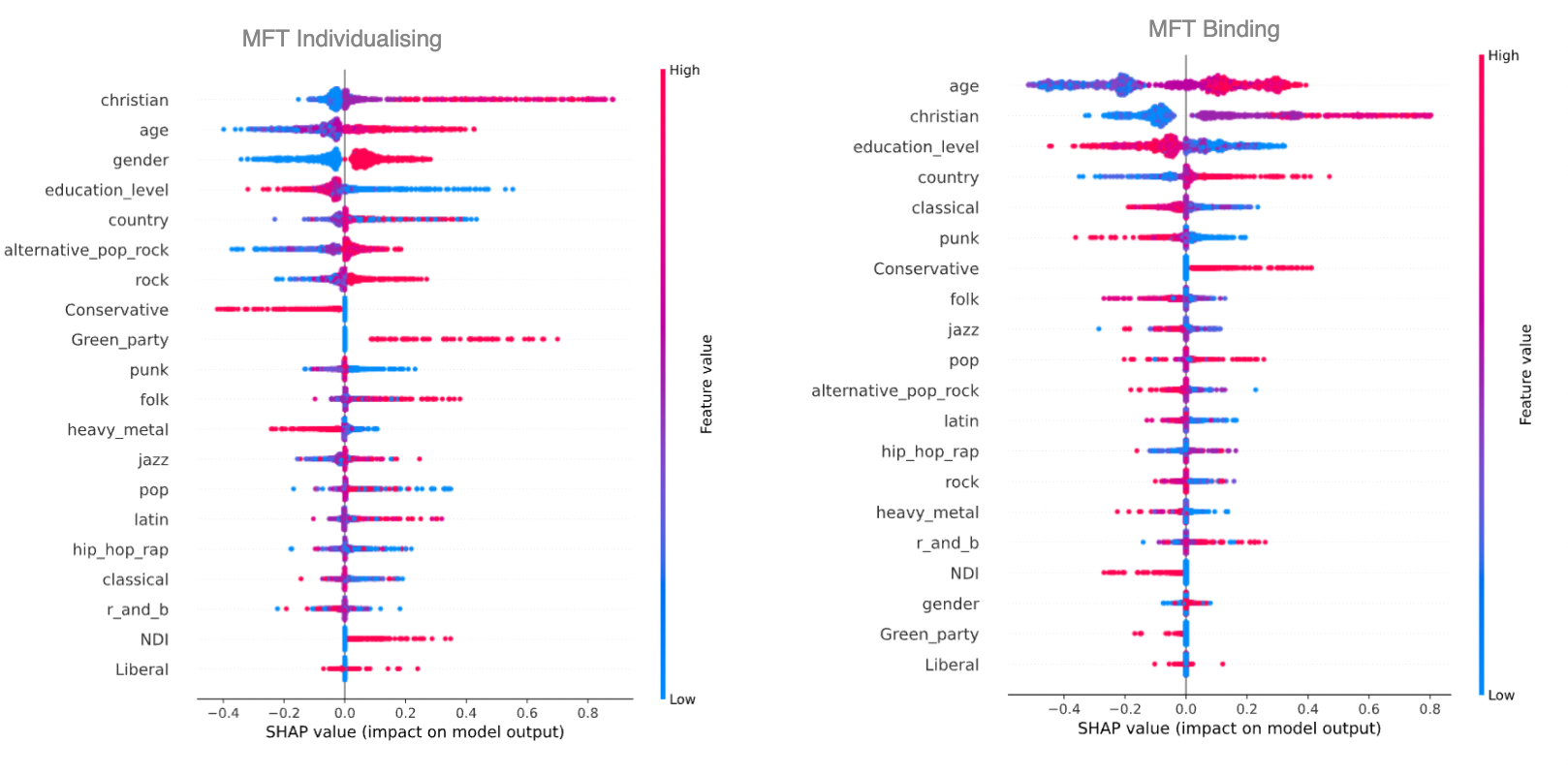}
    \caption{Feature contributions (via SHAP values). The higher the SHAP value (x-axis) the more the feature contributes to the moral prediction.  Horizontal location of the red and blue points reveal whether the value is linked with higher or lower predictions of each moral traits. Whereas the colours (red and blue) show if that variable is high or low for that observation. For example, red points positioned in the left of x-axis, disclose that the variable is negatively correlated with the predicted feature (i.e., Conservatives with  Individualising, etc)}.
    \label{fig:feature_importance}
\end{figure}

Figure \ref{fig:feature_importance} depicts the impact of each feature in the prediction of the two superior foundations, when including demographics, education level, and political views, as well as the 13 music genres (EX6 in Table ~\ref{tab:MFT_predictions_xgb_2}). 
Across all experimental designs, the christian music genre emerged as the most predictive one. Regarding the demographics, the participants' age stands out both for Individualising and Social Binding foundations.

\begin{table}[ht]
\centering
\caption{Moral traits classification with XGBoost for different number of predictors (EX1, EX4, EX5, EX6). Models evaluated based on average weighted AUROC and standard deviation over 5-fold cross-validation (baseline is 50).}
\label{tab:MFT_predictions_xgb_2}
\begin{tabularx}{.8\textwidth}{@{}X C C C C}
\toprule
\multicolumn{5}{c}{Moral Traits Predictors} \\
\midrule               
 & EX1 & EX4 & EX5 & EX6 \\
\midrule
Care & .57 (3.7) & .62 (3.2) & .62 (3.0) & \textbf{.63 (2.3)} \\
Fairness & .56 (2.9) & .58 (2.5) & .57 (2.3) & \textbf{.62 (4.3)} \\
Authority & .63 (0.8) & .64 (1.6) & .65 (2.0) & \textbf{.66 (1.6)} \\
Purity & .69 (2.8) & .71 (3.0) & \textbf{.71 (1.4)} & .71 (1.6) \\
Loyalty & .61 (2.4) & \textbf{.67 (3.5)} & .66 (2.2) & .66 (2.9) \\
\midrule
Individ. & .55 (3.5) & .59 (2.4) & .59 (3.3) & \textbf{.61 (1.8)} \\
Binding & .67 (2.4) & .71 (3.2) & .70 (2.2) &  \textbf{.72 (2.9)}\\
\bottomrule
\end{tabularx}
\end{table}

\subsection{Regression Task}

Data binning is a common way to aggregate information and facilitate the classification tasks. 
However, theare are known issues to dichotomisation of variables which often lead to misleading results
~\cite{maccallum2002practice}.
Here, to ensure that the most predictive features emerged from the classification process, are indeed descriptive of the moral foundation we conducted a regression analysis.
At this point, the aim is to understand whether we can estimate the original moral scores (predicting the quantity) based on our explanatory variables in disposition (i.e. music genres ad demographics). 

To do so, we trained an XGBoost Regressor for each moral foundation. We maintained the same experimental designs and settings as in the classification task (data shuffling, 5-fold cross-validation scheme, with 70\% of the data points used for training and the 30\% for testing).
For model evaluation, we used Mean Absolute Error (MAE). 
These options allow for a direct comparison of the most predictive features with the ones emerged from the classification task.

\begin{table}[ht]
\centering
\caption{Moral traits regression models for the experimental designs EX1, EX2, and EX3, with XGBoost Regression algorithm. Mean Absolute Error (MAE) and standard deviation over 5-fold cross validation was used for the evaluation of each model. Each value on the table measured by MAE represents the average of the absolute error between predicted and actual values, thus, lower MAE scores represent better model performance}
\label{tab:MFT_regression_xgb}
\begin{tabularx}{.8\textwidth}{@{}X C C C }
\toprule
\multicolumn{4}{c}{Music Preference - Regression Task} \\
\midrule
 & EX1 & EX2 & EX3 \\
\midrule
Care & 3.86 (13.2) & \textbf{3.72 (10.9)} & 3.89 (7.0) \\
Fairness & \textbf{3.27 (11.1)} & 3.28 (9.6) & 3.55 (8.7) \\
Authority & \textbf{4.19 (23.3)} & 4.20 (16.7) & 4.47 (13.9) \\
Purity & \textbf{4.86 (19.7)} & 4.99 (25.0) & 5.35 (21.0) \\
Loyalty & 4.46 (12.1) & \textbf{4.33 (19.4)} & 4.64(11.6) \\
\midrule
Individ. & 3.23 (9.5) & \textbf{3.17 (8.5)} & 3.35 (9.9) \\
Binding & 3.86 (15.1) & \textbf{3.79 (6.3)} & 4.22 (18.5)\\
\bottomrule
\end{tabularx}
\end{table}



Table ~\ref{tab:MFT_regression_xgb}, summarises the results for the regression models. 
We notice that as in the classification approach, when adding  information to the models the MAE decreases indicating that the model fits the data better. 
Also in this case the gain of adding more information is relatively small with respect to the music genres alone. 

\begin{figure}[ht]
    \centering
    \includegraphics[scale=0.55]{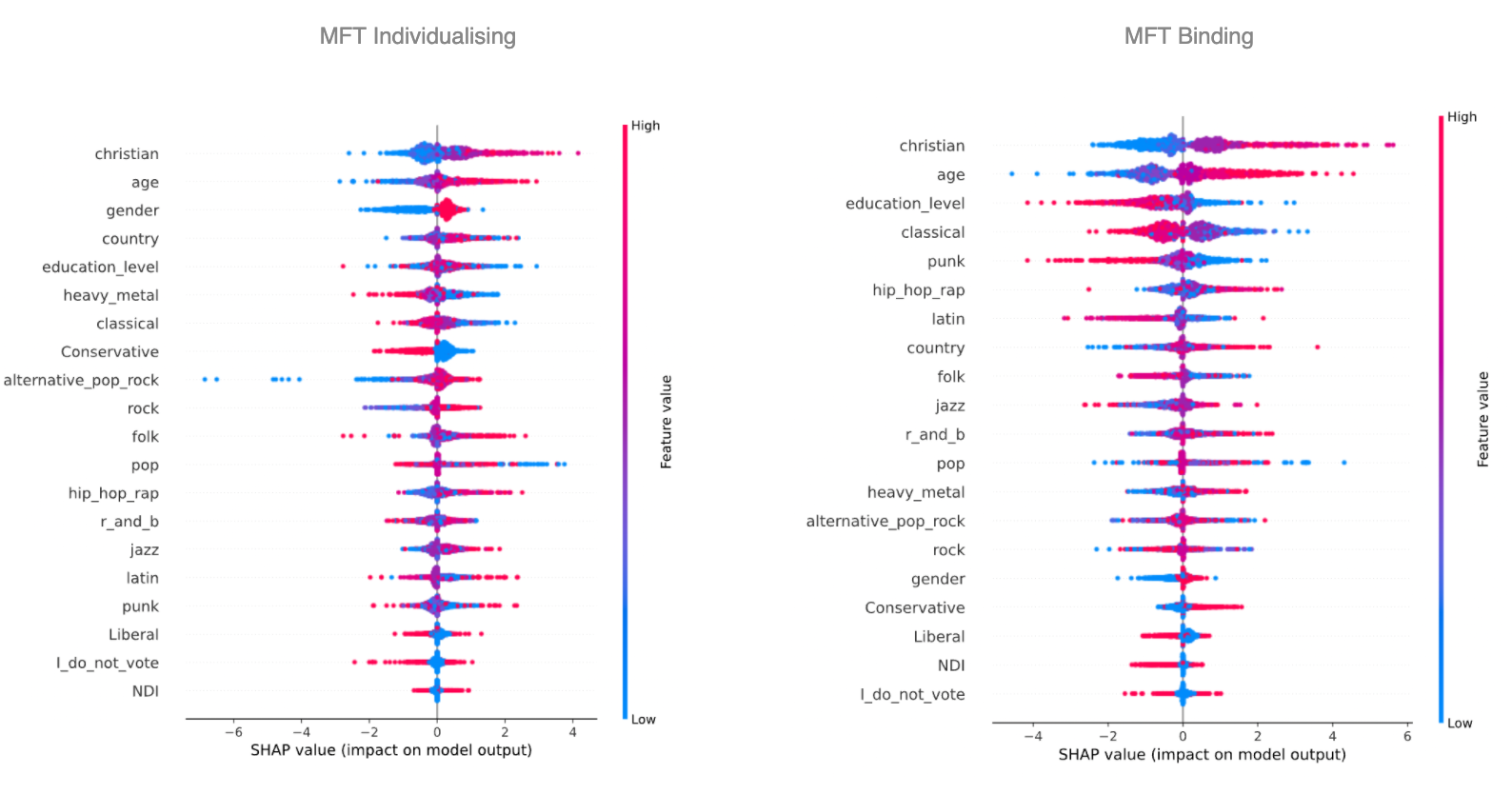}
    \caption{Features with the most impact on the XGBoost regression model output. Horizontal location of the red and blue points reveal whether the value is linked with higher or lower predictions of each moral traits. Whereas the colours (red and blue) show if that variable is high or low for that observation. For example, red points positioned in the left of x-axis, disclose that the variable is negatively correlated with the predicted feature (i.e., punk with MFT Binding, etc).}
    \label{fig:feature_importance_regression}
\end{figure}

We visualise the most predictive features employing again the Shap values (see Figure ~\ref{fig:feature_importance_regression}). Interestingly, the christian music genre appears again as the most important predictor for both the Binding and Individualising foundations while  age is also very important for both models. Generally, the feature importances for the output of the XGboost regressor, is in line with the feature significance obtained with the classification approach. The same holds for all the moral foundations which are not depicted here for spacing issues.



\begin{table}[ht]
\centering
\caption{Moral traits classification with XGBoost for different number of regressors (EX1, EX4, EX5, EX6). Mean Absolute Error (MAE) and standard deviation over 5-fold cross validation was used for the evaluation of each model.}
\label{tab:MFT_regression_xgb_2}
\begin{tabularx}{.8\textwidth}{@{}X C C C C}
\toprule
\multicolumn{5}{c}{Moral Traits Regressors} \\
\midrule
 & EX1 & EX4 & EX5 & EX6 \\
\midrule
Care & 3.86 (13.2) & 3.72 (6.2) & 3.71 (9.3) & \textbf{3.60 (8.0)} \\
Fairness & 3.27 (11.1) & 3.25 (8.2) & 3.19 (10.5) & \textbf{3.12 (13.4)} \\
Authority & 4.19 (23.3) & 4.14 (15.9) & 4.10 (9.3) & \textbf{4.09 (11.0)} \\
Purity & 4.86 (19.7) & 4.86 (20.4) & 4.74 (18.9) & \textbf{4.71 (15.8)} \\
Loyalty & 4.46 (12.1) & \textbf{4.19 (22.8)} & 4.20 (18.7) & 4.21 (18.7) \\
\midrule
Individ. & 3.23 (9.5) & 3.17 (14.4) & 3.17 (8.0) & \textbf{3.0 (8.9)} \\
Binding & 3.86 (15.1) &3.80 (11.0) & 3.76 (13.1) & \textbf{3.74 (5.4)}\\
\bottomrule
\end{tabularx}
\end{table}

\section{Conclusion}

Henry Wadsworth Longfellow wrote, ``Music is the universal language of mankind.''   
Contemporary research has found converging evidence that people listen to music that reflects their psychological traits and needs and help express emotions, cultures, values and personalities.

In this paper, we analysed the less explored links between musical preferences, demographics (age, gender, political views, and education level) and Moral Foundations (MFT \cite{Graham2011}). We applied both classification and regression models for moral traits prediction. From classification results, it was inferred that MFT Binding (encompassing purity, authority and loyalty) was best predicted with AUROC score 72\%, whereas MFT individualising (including care and fairness) showed weaker results with AUROC score 61\%. While, for the regression task the lowest MAE was 3.0 for the  Individualising and 3.74 for Social Binding. In both approaches, the most impactful features on inferring morality were christian music and age. 

Moral foundations are strongly tied to political views; despite that, the musical features are more predictive than political leanings.
Social binding is related to conservative political views \cite{graham2009liberals} - and in fact is predicted by christian, and country music. 
We notice that people instinctually express their moral values through the music they listen to. 
We instinctively  \textit{categorize} objects, symbols, but also people,  
creating a notion of \textit{social identity}.
According to the social identity theory members of a group will seek to find negative aspects to other groups thus enhancing their self-image~\cite{tajfel1979integrative}.
Such reasoning reflects on a broad range of attitudes related to stereotype formations~\cite{miller2010self} but also as we notice here to musical preferences.
For instance, people higher in social binding foundations (authority, loyalty, and purity) tend to listen to country music which often expresses notions of patriotism. Christian music is also a predictor of this superior foundation, which again fosters the notion of belonging to a group. On the other hand, music genres such as punk, and hip hop are know to challenge the traditional values and the status quo, hence are preferred by people who strongly value these aspects.

To conclude, our findings suggested that musical preferences are quite informative of deeper psychological attributes; still there is space for improvement. For instance, we noticed that the care, fairness, and loyalty foundations are very hard to predict; to this extend we aim to explore  musical content analysis; incorporating linguistic cues as well as the moral valence scores proposed by Araque et al. \cite{araque2020moralstrength} on lyrics will improve the performance. 

Generally,  self-reported preferences for music genres reflect, at least partially, preferences for external properties of music. We aim to further investigate the association between music listening preferences and moral traits and the link of music with other psychological aspects such as human values and emotions. 
In future work we aim to delve deeper into the relation between music and morality, and between music and other universal human values, by using passively collected digital traits of music listening behaviours outside a laboratory setting and over a period of time \cite{anderson2020just}.
Developing data-informed models will help unlock the potential of personalised, uniquely tailored digital music experiences and communication strategies \cite{kalimeri2019predicting,anderson2020algorithmic}. Predicting the moral values from listening behaviours can provide noninvasive insights on the values or other psychological aspects of populations at a large scale.


\begin{thebibliography}{10}

\bibitem{anderson2020algorithmic}
Ashton Anderson, Lucas Maystre, Ian Anderson, Rishabh Mehrotra, and Mounia
  Lalmas.
\newblock Algorithmic effects on the diversity of consumption on spotify.
\newblock In {\em Proceedings of The Web Conference 2020}, pages 2155--2165,
  2020.

\bibitem{anderson2020just}
Ian Anderson, Santiago Gil, Clay Gibson, Scott Wolf, Will Shapiro, Oguz
  Semerci, and David~M Greenberg.
\newblock ``just the way you are'': Linking music listening on spotify and
  personality.
\newblock {\em Social Psychological and Personality Science}, page
  1948550620923228, 2020.

\bibitem{ansani2019you}
Alessandro Ansani, Francesca D'Errico, and Isabella Poggi.
\newblock `you will be judged by the music i hear': A study on the influence of
  music on moral judgement.
\newblock In {\em Web Intelligence}, volume~17, pages 53--62. IOS Press, 2019.

\bibitem{araque2020moralstrength}
Oscar Araque, Lorenzo Gatti, and Kyriaki Kalimeri.
\newblock Moralstrength: Exploiting a moral lexicon and embedding similarity
  for moral foundations prediction.
\newblock {\em Knowledge-based systems}, 191:105184, 2020.

\bibitem{carr2012group}
Catherine Carr, Patricia d'Ardenne, Ann Sloboda, Carleen Scott, Duolao Wang,
  and Stefan Priebe.
\newblock Group music therapy for patients with persistent post-traumatic
  stress disorder--an exploratory randomized controlled trial with mixed
  methods evaluation.
\newblock {\em Psychology and Psychotherapy: Theory, Research and Practice},
  85(2):179--202, 2012.

\bibitem{chen2016xgboost}
Tianqi Chen and Carlos Guestrin.
\newblock Xgboost: A scalable tree boosting system.
\newblock In {\em Proceedings of the 22nd acm sigkdd international conference
  on knowledge discovery and data mining}, pages 785--794, 2016.

\bibitem{cross2009evolution}
Ian Cross and Iain Morley.
\newblock The evolution of music: Theories, definitions and the nature of the
  evidence.
\newblock {\em Communicative musicality: Exploring the basis of human
  companionship}, pages 61--81, 2009.

\bibitem{devenport2019predicting}
Scott~P Devenport and Adrian~C North.
\newblock Predicting musical taste: Relationships with personality aspects and
  political orientation.
\newblock {\em Psychology of Music}, 47(6):834--847, 2019.

\bibitem{gardikiotis2012rock}
Antonis Gardikiotis and Alexandros Baltzis.
\newblock `rock music for myself and justice to the world!': Musical identity,
  values, and music preferences.
\newblock {\em Psychology of Music}, 40(2):143--163, 2012.

\bibitem{gold2009dose}
Christian Gold, Hans~Petter Solli, Viggo Kr{\"u}ger, and Stein~Atle Lie.
\newblock Dose--response relationship in music therapy for people with serious
  mental disorders: Systematic review and meta-analysis.
\newblock {\em Clinical psychology review}, 29(3):193--207, 2009.

\bibitem{graham2009liberals}
Jesse Graham, Jonathan Haidt, and Brian~A Nosek.
\newblock Liberals and conservatives rely on different sets of moral
  foundations.
\newblock {\em Journal of personality and social psychology}, 96(5):1029, 2009.

\bibitem{Graham2011}
Jesse Graham, Brian Nosek, Jonathan Haidt, Ravi Iyer, Spassena Koleva, and
  Peter~H Ditto.
\newblock {Mapping the moral domain.}
\newblock {\em Journal of personality and social psychology}, 101(2):366--85,
  August 2011.

\bibitem{greenberg2015personality}
David~M Greenberg, Daniel M{\"u}llensiefen, Michael~E Lamb, and Peter~J
  Rentfrow.
\newblock Personality predicts musical sophistication.
\newblock {\em Journal of Research in Personality}, 58:154--158, 2015.

\bibitem{haidt2004intuitive}
Jonathan Haidt and Craig Joseph.
\newblock Intuitive ethics: How innately prepared intuitions generate
  culturally variable virtues.
\newblock {\em Daedalus}, 133(4):55--66, 2004.

\bibitem{hole2015music}
Jenny Hole, Martin Hirsch, Elizabeth Ball, and Catherine Meads.
\newblock Music as an aid for postoperative recovery in adults: a systematic
  review and meta-analysis.
\newblock {\em The Lancet}, 386(10004):1659--1671, 2015.

\bibitem{kalimeri2019predicting}
Kyriaki Kalimeri, Mariano~G Beir{\'o}, Matteo Delfino, Robert Raleigh, and Ciro
  Cattuto.
\newblock Predicting demographics, moral foundations, and human values from
  digital behaviours.
\newblock {\em Computers in Human Behavior}, 92:428--445, 2019.

\bibitem{kalimeri2019human}
Kyriaki Kalimeri, Mariano G.~Beir{\'o}, Alessandra Urbinati, Andrea Bonanomi,
  Alessandro Rosina, and Ciro Cattuto.
\newblock Human values and attitudes towards vaccination in social media.
\newblock In {\em Companion Proceedings of The 2019 World Wide Web Conference},
  pages 248--254, 2019.

\bibitem{kim2013moral}
Eunkyung Kim, Ravi Iyer, Jesse Graham, Yu-Han Chang, and Rajiv Maheswaran.
\newblock Moral values from simple game play.
\newblock In {\em International Conference on Social Computing,
  Behavioral-Cultural Modeling, and Prediction}, pages 56--64. Springer, 2013.

\bibitem{krismayer2019predicting}
Thomas Krismayer, Markus Schedl, Peter Knees, and Rick Rabiser.
\newblock Predicting user demographics from music listening information.
\newblock {\em Multimedia Tools and Applications}, 78(3):2897--2920, 2019.

\bibitem{lin2011mental}
Shuai-Ting Lin, Pinchen Yang, Chien-Yu Lai, Yu-Yun Su, Yi-Chun Yeh, Mei-Feng
  Huang, and Cheng-Chung Chen.
\newblock Mental health implications of music: Insight from neuroscientific and
  clinical studies.
\newblock {\em Harvard review of psychiatry}, 19(1):34--46, 2011.

\bibitem{loersch2013unraveling}
Chris Loersch and Nathan~L Arbuckle.
\newblock Unraveling the mystery of music: Music as an evolved group process.
\newblock {\em Journal of Personality and Social Psychology}, 105(5):777, 2013.

\bibitem{lonsdale2011we}
Adam~J Lonsdale and Adrian~C North.
\newblock Why do we listen to music? a uses and gratifications analysis.
\newblock {\em British journal of psychology}, 102(1):108--134, 2011.

\bibitem{lundberg2017unified}
Scott Lundberg and Su-In Lee.
\newblock A unified approach to interpreting model predictions.
\newblock {\em arXiv preprint arXiv:1705.07874}, 2017.

\bibitem{lundberg2019explainable}
Scott~M. Lundberg, Gabriel Erion, Hugh Chen, Alex DeGrave, Jordan~M. Prutkin,
  Bala Nair, Ronit Katz, Jonathan Himmelfarb, Nisha Bansal, and Su-In Lee.
\newblock Explainable ai for trees: From local explanations to global
  understanding, 2019.

\bibitem{maccallum2002practice}
Robert~C MacCallum, Shaobo Zhang, Kristopher~J Preacher, and Derek~D Rucker.
\newblock On the practice of dichotomization of quantitative variables.
\newblock {\em Psychological methods}, 7(1):19, 2002.

\bibitem{macdonald2013music}
Raymond MacDonald, Gunter Kreutz, and Laura Mitchell.
\newblock {\em Music, health, and wellbeing}.
\newblock Oxford University Press, 2013.

\bibitem{mcadams2006new}
Dan~P McAdams and Jennifer~L Pals.
\newblock A new big five: fundamental principles for an integrative science of
  personality.
\newblock {\em American psychologist}, 61(3):204, 2006.

\bibitem{miller2010self}
Saul~L Miller, Jon~K Maner, and D~Vaughn Becker.
\newblock Self-protective biases in group categorization: Threat cues shape the
  psychological boundary between ``us'' and ``them''.
\newblock {\em Journal of personality and social psychology}, 99(1):62, 2010.

\bibitem{nave2018musical}
Gideon Nave, Juri Minxha, David~M Greenberg, Michal Kosinski, David Stillwell,
  and Jason Rentfrow.
\newblock Musical preferences predict personality: evidence from active
  listening and facebook likes.
\newblock {\em Psychological Science}, 29(7):1145--1158, 2018.

\bibitem{north2004uses}
Adrian~C North, David~J Hargreaves, and Jon~J Hargreaves.
\newblock Uses of music in everyday life.
\newblock {\em Music perception}, 22(1):41--77, 2004.

\bibitem{tajfel1979integrative}
Henri Tajfel, John~C Turner, William~G Austin, and Stephen Worchel.
\newblock An integrative theory of intergroup conflict.
\newblock {\em Organizational identity: A reader}, 56(65):9780203505984--16,
  1979.

\end{thebibliography}

\end{document}